\documentclass[twocolumn,structabstract]{aa}

\usepackage{txfonts}
\usepackage{natbib}
\bibpunct{(}{)}{;}{a}{}{,}
\usepackage{graphicx}

\begin{document}

\title{Three-dimensional magnetohydrodynamic simulation
  of the solar magnetic flux emergence}
\subtitle{Parametric study on the horizontal divergent flow}
\author{Shin~Toriumi \and Takaaki~Yokoyama}
\institute{Department of Earth and Planetary Science,
  University of Tokyo,
  7-3-1 Hongo, Bunkyo-ku, Tokyo 113-0033, Japan\\
  \email{toriumi@eps.s.u-tokyo.ac.jp}
}
\date{Received / Accepted}

\abstract
{Solar active regions are formed
  through the emergence of magnetic flux
  from the deeper convection zone.
  Recent satellite observations have shown that
  a horizontal divergent flow (HDF) stretches out over
  the solar surface
  just before the magnetic flux appearance.}
{The aims of this study
  are to investigate the driver of the HDF
  and to see the dependency
  of the HDF
  on the parameters
  of the magnetic flux
  in the convection zone.}
{We conduct three-dimensional magnetohydrodynamic
  (3D MHD) numerical simulations
  of the magnetic flux emergence
  and vary the parameters
  in the initial conditions.
  An analytical approach is also taken
  to explain the dependency.}
{The horizontal gas pressure gradient
  is found to be 
  the main driver of the HDF.
  The maximum HDF speed shows
  positive correlations with
  the field strength and twist intensity.
  The HDF duration has
  a weak relation with the twist,
  while it shows negative dependency
  on the field strength
  only in the case of the stronger field regime.}
{Parametric dependencies analyzed in this study
  may allow us to probe the structure
  of the subsurface magnetic flux
  by observing properties of the HDF.}


\authorrunning{S.~Toriumi \& T.~Yokoyama}
\titlerunning{3D MHD simulation on the solar flux emergence}

\maketitle

\section{Introduction}

The dynamics of the rising magnetic flux
and the formation process
of solar active regions
have widely been investigated
through a series of
analytical and numerical studies.
\citet{par75} first calculated
the rising speed of a flux tube
in the convection zone (CZ)
by considering a force balance
between the magnetic buoyancy
and the aerodynamic drag
acting on the flux tube,
while \citet{sch79} simulated
a buoyant emergence
of a flux tube in the CZ
in a two-dimensional (2D) scheme.
\citet{mor96} and \citet{emo98}
investigated the dependency of the rising tube
on the initial twist intensity
and found that the tube needs
a certain degree of twist
to hold its coherency.
After \citet{shi89} and \citet{mag01}
conducted 2D simulations of flux emergence
from the surface layer
into the corona,
3D simulations have been carried out
by, among others,
\citet{fan01} and \citet{arc04}.
\citet{mur06} simulated
other 3D flux emergences
of this kind
and surveyed the tube's dependency
on the initial field strength
and twist intensity.

Recently,
\citet{tor10,tor11,tor12a}
combined the CZ, the photosphere/chromosphere,
and the corona
into a single computational domain
and simulated magnetic flux emergence
from a deeper CZ both in 2D and 3D.
As a result,
the initial flux
placed at a depth of $-20\ {\rm Mm}$
starts its emergence
in the solar interior,
which then slows down gradually
in the uppermost CZ.
This is because
the plasma pushed up
by the emerging flux
rises to the isothermally-stratified
(i.e. convectively-stable)
surface layer
and is then trapped and compressed
between them,
which, in turn,
suppresses the rising flux
from below.
Such compressed plasma
will escape laterally
around the photospheric layer
from the rising flux
as a horizontal divergent flow (HDF),
just before the flux itself
reaches the surface.
Using SDO/HMI data,
\citet{tor12b} observed
the emerging active region
located away from the solar disk center
and found the HDF
in the Dopplergram,
up to about 100 min
before the start
of the flux emergence.

In the present study,
we report
the results of the parametric survey
of the 3D magnetohydrodynamic (MHD)
flux emergence simulation.
The aims of this study are
to investigate which force drives the HDF
and to observe
the dependence of the HDF
on the parameters in the simulation.
One important feature of this HDF study
is that it can be a probe
for exploring the physical state of
the magnetic field
in the upper CZ.
That is,
we may be able to obtain
valuable information
on the subsurface layers
from the direct optical observation
at the surface.
Therefore, in this numerical study,
we vary the parameters of the initial flux tube,
and then check the characteristics
of the consequent HDF
seen at the surface layer.

In the next section
we introduce the basic setup
of the numerical calculation.
In Section \ref{sec:results}
we show the results of the parametric survey,
and in Section \ref{sec:analytic}
we provide some analytic explanations
of the results.
We finally summarize the paper
in Section \ref{sec:summary}.

\section{Numerical Setup
 \label{sec:setup}}

The basic MHD equations,
normalizing units,
computational domain size,
grid spacings,
boundary conditions,
and background stratification
are the same as those
in \citet{tor12a}.
The MHD equations
in vector form are:
\begin{eqnarray}
  \frac{\partial\rho}{\partial t}
  + \mbox{\boldmath $\nabla$}
    \cdot(\rho\mbox{\boldmath $V$})=0,
\end{eqnarray}
\begin{eqnarray}
  \frac{\partial}{\partial t}
  (\rho\mbox{\boldmath $V$})
  + \mbox{\boldmath $\nabla$}\cdot
  \left(
    \rho\mbox{\boldmath $V$}\mbox{\boldmath $V$}
    +p\mbox{\boldmath $I$}
    -\frac{\mbox{\boldmath $BB$}}{4\pi}
    +\frac{\mbox{\boldmath $B$}^{2}}{8\pi}\mbox{\boldmath $I$}
  \right)
  -\rho\mbox{\boldmath $g$}=0,
\end{eqnarray}
\begin{eqnarray}
  \frac{\partial\mbox{\boldmath $B$}}
       {\partial t}
  = \mbox{\boldmath $\nabla$}
    \times (\mbox{\boldmath $V$}\times\mbox{\boldmath $B$}),
\end{eqnarray}
\begin{eqnarray}
  &&\frac{\partial}{\partial t}
  \left(
    \rho U
    + \frac{1}{2}\rho\mbox{\boldmath $V$}^{2}
    + \frac{\mbox{\boldmath $B$}^{2}}{8\pi}
  \right) \nonumber \\
  &&+\mbox{\boldmath $\nabla$}\cdot
  \left[
    \left(
      \rho U + p + \frac{1}{2}\rho\mbox{\boldmath $V$}^{2}
    \right)
    \mbox{\boldmath $V$}
    + \frac{c}{4\pi}
      \mbox{\boldmath $E$}\times\mbox{\boldmath $B$}
  \right]
  -\rho\mbox{\boldmath $g$}\cdot\mbox{\boldmath $V$}
  =0,
\end{eqnarray}
and
\begin{eqnarray}
  U=\frac{1}{\gamma-1}
    \frac{p}{\rho},
\end{eqnarray}
\begin{eqnarray}
  \mbox{\boldmath $E$}
  =-\frac{1}{c}
   \mbox{\boldmath $V$}\times\mbox{\boldmath $B$},
\end{eqnarray}
\begin{eqnarray}
  p=\frac{k_{\rm B}}{m}\rho T,
\end{eqnarray}
where $\rho$ denotes the gas density,
$\mbox{\boldmath $V$}$ velocity vector,
$p$ pressure,
$\mbox{\boldmath $B$}$ magnetic field,
$c$ the speed of light,
$\mbox{\boldmath $E$}$ electric field,
and $T$ temperature,
while $U$ is the internal energy per unit mass,
$\mbox{\boldmath $I$}$
the unit tensor,
$k_{\rm B}$ the Boltzmann constant,
$m (={\rm const.})$ the mean molecular mass,
and $\mbox{\boldmath $g$}$
the uniform
gravitational acceleration.
We assume the medium to be
an inviscid perfect gas
with a specific heat ratio
$\gamma =5/3$.
All the physical values
are normalized by
the pressure scale height
$H_{0}=200\ {\rm km}$ for length,
the sound speed $C_{\rm s0}=8\ {\rm km\ s}^{-1}$ for velocity,
$\tau_{0}\equiv H_{0}/C_{\rm s0}=25\ {\rm s}$ for time,
and $\rho_{0}=1.4\times 10^{-7}\ {\rm g\ cm}^{-3}$ for density,
all of which 
are the typical values
in the photosphere.
The units for pressure, temperature,
and magnetic field strength are
$p_{0}=9.0\times 10^{4}\ {\rm dyn\ cm}^{-2}$,
$T_{0}=4000\ {\rm K}$,
and $B_{0}=300\ {\rm G}$, respectively.

Here, 3D Cartesian coordinates
$(x, y, z)$ are used,
where $z$ is parallel to the gravitational acceleration vector,
$\mbox{\boldmath $g$}=(0,0,-g_{0})$,
and $g_{0}=C_{\rm s0}^{2}/(\gamma H_{0})$
by definition. 
The simulation domain is
$(-400,-200,-200)\leq (x/H_{0},y/H_{0},z/H_{0})\leq (400,200,250)$,
resolved by
$1602\times 256\times 1024$ grids.
In the $x$-direction,
the mesh size is
$\Delta x/H_{0}=0.5$ (uniform).
In the $y$-direction ($z$-direction),
the mesh size is
$\Delta y/H_{0}=0.5$
($\Delta z/H_{0}=0.2$)
in the central area of the domain,
which gradually increases
for each direction.
We assume periodic boundaries
for both horizontal directions
and symmetric boundaries
for the vertical direction.

The background atmosphere
consists of three different layers.
From the bottom,
the layers are
the adiabatically stratified CZ,
the cool isothermal
photosphere/chromosphere,
and the hot isothermal corona.
The stratification
in the CZ ($z/H_{0}<0$)
is given as
\begin{eqnarray}
  T=T_{\rm ph}-z
   \left|
     \frac{dT}{dz}
   \right|_{\rm ad},
\end{eqnarray}
where $T_{\rm ph}/T_{0}=1$
is the respective temperature
in the photosphere/chromosphere
and
\begin{eqnarray}
  \left| \frac{dT}{dz} \right|_{\rm ad}
  = \frac{\gamma-1}{\gamma}
    \frac{mg_{0}}{k_{\rm B}}
\end{eqnarray}
is the adiabatic temperature gradient.
The profile above the surface is
\begin{eqnarray}
  T(z)=T_{\rm ph}+\frac{1}{2}(T_{\rm cor}-T_{\rm ph})
  \left\{
    \tanh{ \left[
        \frac{z-z_{\rm cor}}{w_{\rm tr}}
      \right]
      +1 }
  \right\},
\end{eqnarray}
where $T_{\rm cor}/T_{0}=100$
is the temperatures
in the corona,
$z_{\rm cor}/H_{0}=10$
is the base of the corona,
and $w_{\rm tr}/H_{0}=0.5$
is the transition scale length.
Based on the temperature distribution above,
the pressure and density profiles
are defined
by the equation of static pressure balance:
\begin{eqnarray}
  \frac{dp(z)}{dz} + \rho(z) g_{0} =0.
\end{eqnarray}

The initial flux tube is embedded
in the CZ
at $z_{\rm tube}/H_{0}=-100$,
i.e., $z_{\rm tube}=-20\ {\rm Mm}$,
of which the axial and azimuthal profiles
are given as
\begin{eqnarray}
  \left\{
    \begin{array}{lll}
      B_{x}(r) &=&
      B_{\rm tube}\exp{\left( -\displaystyle\frac{r^{2}}{R_{\rm tube}^{2}} \right)}\\
      B_{\phi}(r) &=& qrB_{x}(r)
    \end{array}
    \right.,
    \label{eq:tube}
\end{eqnarray}
respectively,
where $B_{\rm tube}$ is
the axial field strength,
$r$ the radial distance
from the tube's center
$(y_{\rm tube}/H_{0},z_{\rm tube}/H_{0})=(0,-100)$,
$R_{\rm tube}$ the typical radial size,
and $q$ the twist intensity.
For the pressure balance
between the field and the plasma,
the pressure distribution inside the tube
is defined as
$p_{\rm i}= p(z) + \delta p_{\rm exc}$
(the subscript ``i''
denotes inside the tube),
where the pressure excess
$\delta p_{\rm exc} (<0)$
is described as
\begin{eqnarray}
  \delta p_{\rm exc}
  = \frac{B_{x}^{2}(r)}{8\pi}
  \left[
    q^{2}
    \left(
      \frac{R_{\rm tube}^{2}}{2}
      -r^{2}
    \right)
    -1
  \right].
\end{eqnarray}
The density inside the tube
is also defined as
$\rho_{\rm i}=\rho(z) + \delta \rho_{\rm exc}$,
where
\begin{eqnarray}
  \delta \rho_{\rm exc}
  =  \rho(z)
  \frac{\delta p_{\rm exc}}{p(z)}
  \exp{
    \left(
      -\frac{x^{2}}{\lambda^{2}}
    \right)
  },
\end{eqnarray}
and $\lambda$ is the perturbation wavelength.
That is,
the middle of the tube,
$x/H_{0}=0$,
is in thermal equilibrium
with external media
and is most buoyant.
The buoyancy decreases
as $|x|/H_{0}$ increases.

\begin{table}
  \caption{Summary of the simulation cases
    \label{tab:parameter}}
  \centering
  \begin{tabular}{cccc}
    \hline\hline
    Case\tablefootmark{a} & Field strength $B_{\rm tube}/B_{0}$ &
    Twist $qH_{0}$ & Wavelength $\lambda/H_{0}$ \\
    \hline
    A & $67$ & $0.1$ & $400$\\
    B & $133$ & $0.1$ & $400$\\
    C & $33$ & $0.1$ & $400$\\
    D & $67$ & $0.2$ & $400$\\
    E & $67$ & $0.05$ & $400$\\
    F & $67$ & $0.1$ & $100$\\
    G & $67$ & $0.1$ & $25$\\
    \hline
  \end{tabular}
  \tablefoot{
    \tablefoottext{a}{Case A is the same as
      that simulated in \citet{tor12a}.
      Cases B and C are for different field strengths
      than that of case A,
      while D and E are for different twists,
      and F and G different wavelengths.}
  }
\end{table}

The parameters we varied
are the field strength
$B_{\rm tube}$,
the twist $q$,
and the perturbation wavelength
$\lambda$.
Table \ref{tab:parameter}
summarizes the cases
in this study.
The case simulated
in \citet{tor12a}
is named here as case A,
while cases B--F are
for different field strength,
twist, and wavelength
than those of A.
Here we fixed the tube's radial size
at $R_{\rm tube}/H_{0}=5$
for all the cases.
It should be noted that
the critical twist
for the kink instability
is $qH_{0}=0.2$
\citep{lin96}.
Therefore,
all the tubes examined here
are stable or,
at least, marginally stable
against the instability
at the beginning
of the calculation.

\section{Simulation Results
  \label{sec:results} }

\subsection{General Evolution
  \label{sec:general}}

\begin{figure}
  \centering
  \includegraphics[scale=1.,clip]{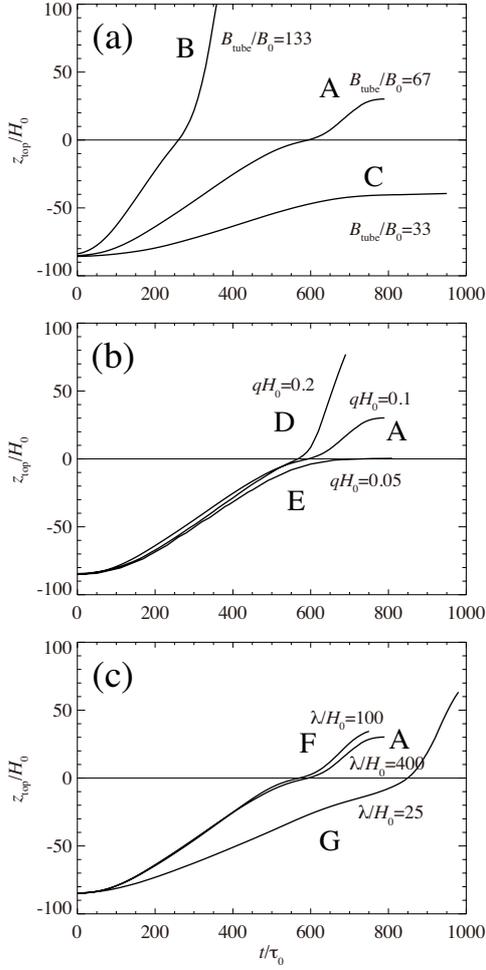}
  \caption{Height-time evolution
    of the flux tube.
    (a) Cases for different $B_{\rm tube}$.
    (b) Cases for different $q$.
    (c) Cases for different $\lambda$.}
  \label{fig:results}
\end{figure}

\begin{figure}
  \centering
  \includegraphics[scale=1.,clip]{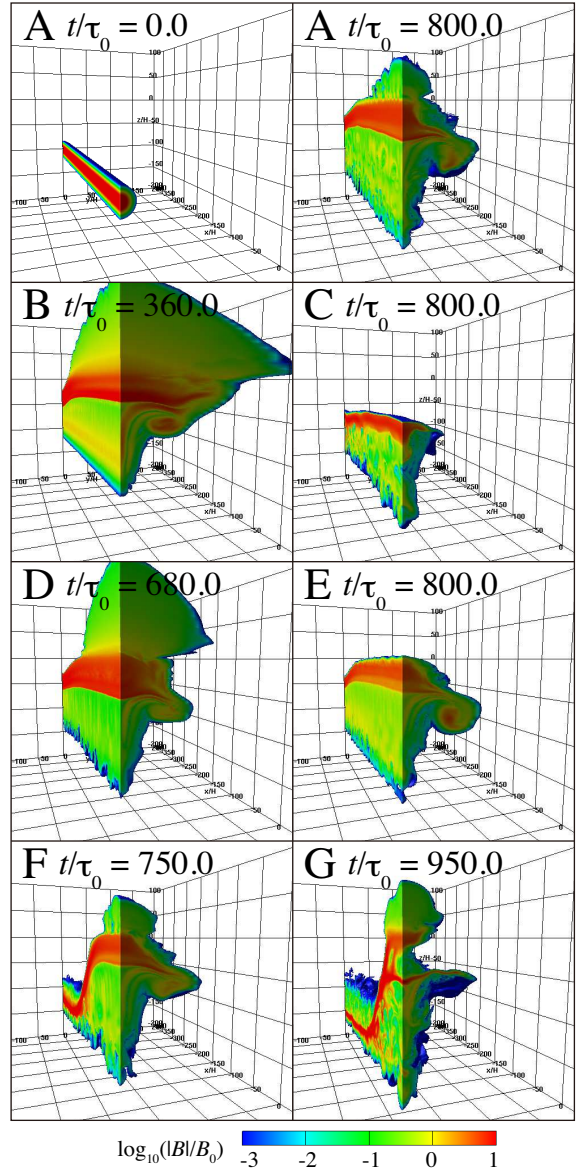}
  \caption{Total magnetic field strength
    of the initial condition for case A
    and the final states for cases A to G,
    plotted over the range
    $-400\le x/H_{0}\le 0$
    and $0\le y/H_{0}\le 200$.}
  \label{fig:avs}
\end{figure}

Fig. \ref{fig:results}
shows the temporal evolution
of the apex of the rising tube,
$z_{\rm top}(t)$.
Also, in Fig. \ref{fig:avs},
we plot the total field strength,
$\log_{10}{(|B|/B_{0})}$,
of the initial condition for case A
and the final states
for all the cases.
As can be seen
in Fig. \ref{fig:results}a,
it is clear that
the tubes with stronger field $B_{\rm tube}$
rise faster.
The rising speed of each tube
in the CZ
is in simple proportion
to the initial field strength,
which is well in accordance with
\citet{mur06} and other previous studies.
Case A,
which has a middle field strength,
shows deceleration
just before it reaches the surface.
This deceleration is the result
of the plasma accumulation,
which is caused by the trapping of material
between the rising tube
and the isothermally-stratified photosphere
above the tube.
After a while,
the tube then starts
further emergence
into the atmosphere
(see the top panels
in Fig. \ref{fig:avs}).
As for the strongest case in B,
the accumulation becomes
less marked,
and thus the tube almost directly passes
through the surface layer
and expands into the higher corona,
without undergoing strong deceleration
(Fig. \ref{fig:avs}B).
When the field is very weak,
as in case C,
the tube stops its emergence
halfway to the surface,
since the tube's buoyancy
is not strong enough
to continue its emergence
(Fig. \ref{fig:avs}C).

Fig. \ref{fig:results}b
shows
the evolution with different twist $q$.
In this figure,
all three tubes
are seen to rise
almost at the same rate
in the CZ,
which is again consistent
with previous studies
\citep[e.g.][]{mur06}.
When the twist is weak
and thus cannot hold the coherency
(case E),
the tube expands
and suffers
deformation
by aerodynamic drag.
As a result,
the tube cannot maintain
a strong enough magnetic field
to continue its further emergence
(Fig. \ref{fig:avs}E).

Fig. \ref{fig:results}c
compares three cases with
different wavelengths
of the initial perturbation.
The initial wavelength
is crucial for two factors:
the curvature force (magnetic tension),
which pulls down the rising tube,
and the drainage of the internal media
due to gravity,
which encourages the emergence.
When the wavelength is smaller,
the curvature force is expected to be stronger,
while the drainage becomes more effective.
In Fig. \ref{fig:results}c
the rising velocities
of cases A and F
are almost the same,
which indicates that
both effects
cancel each other out.
However,
the shortest wavelength tube
(case G)
shows a much slower emergence rate
in the CZ,
which indicates that
the curvature force is more effective
and slows down the emergence.
As for the emergence above the surface layer,
on the contrary,
Fig. \ref{fig:results}c shows
an exactly opposite trend that
the rising is much faster
when the wavelength is shortest
(case G).
One may find that,
in Fig \ref{fig:avs}G,
the main tube remains
in the CZ at around $z/H_{0}=-50$,
while the upper part has been detached
from the main tube
and has started further emergence
into the atmosphere.
The reason for the rapid ascent
may be because,
in the shortest wavelength case,
namely, in the highly curved loops,
the draining of the plasma
from the apex is more effective,
which helps the faster emergence
above the photosphere.

\subsection{Driver of the HDF}

\begin{figure}
  \centering
  \includegraphics[scale=1.1,clip]{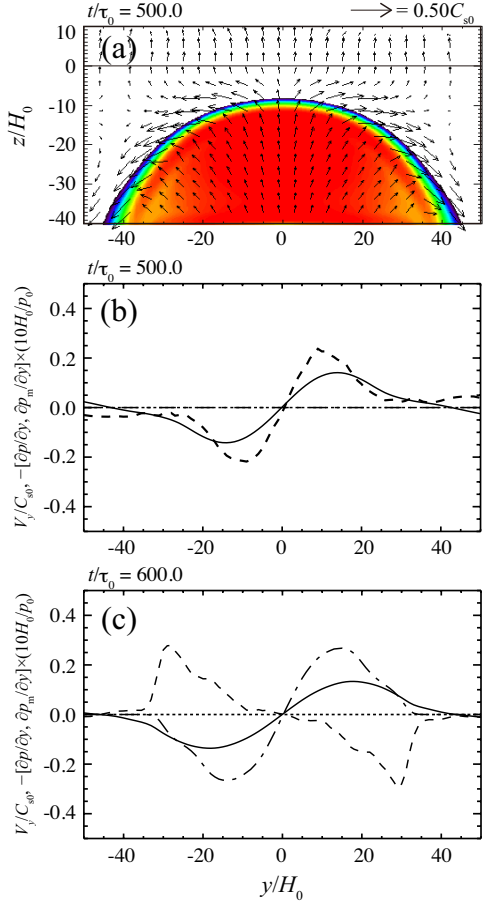}
  \caption{(a) Cross-sectional profile
    of the rising magnetic flux tube
    (case A).
    Plotted value is
    the logarithmic field strength
    $\log_{10}{(|B|/B_{0})}$
    averaged over $6.5\le x/H_{0}\le 13.5$,
    at the time $t/\tau_{0}=500$.
    The color saturates at
    $|B|/B_{0}=1.0$ (red)
    and $-4.0$ (purple).
    (b) Horizontal velocity $V_{y}/C_{\rm s0}$
    (thick solid),
    pressure gradient
    $-\partial p/\partial y\times(10H_{0}/p_{0})$
    (dashed),
    and magnetic pressure gradient
    $-\partial p_{\rm m}/\partial y\times(10H_{0}/p_{0})$
    (dash-dotted),
    at $t/\tau_{0}=500$.
    (c) Same as (b) but for
    $t/\tau_{0}=600$.
  }
  \label{fig:driver}
\end{figure}

Fig. \ref{fig:driver}a
is the cross-sectional distribution
of the field strength
of Case A.
The plotted value is
the logarithmic field strength
$\log_{10}{(|B|/B_{0})}$
averaged over
$6.5\le x/H_{0}\le 13.5$,
i.e., around $x/H_{0}=10$.
The reason we choose this $x$-range
is to select one folded structure
at the tube's surface
\citep[see Fig. 3 of][]{tor12a}.
In this figure,
there is a flow field
in front of the rising flux tube,
which is flowing
from the apex to the flanks
of the tube.
One characteristic of this plasma layer
is the horizontal divergent flow (HDF)
that is seen
at the solar surface
just before the flux tube itself
emerges.

To investigate
which force drives the HDF,
in Figs. \ref{fig:driver}b and c,
we plot the horizontal flow velocity
$V_{y}$,
pressure gradient
$-\partial p/\partial y$,
and magnetic pressure gradient
$-\partial p_{\rm m}/\partial y$,
averaged over $6.5\le x/H_{0}\le 13.5$
and $-10\le z/H_{0}\le 0$,
where $p_{\rm m}=B^{2}/(8\pi)$.
Magnetic tension is not plotted here,
since it is rather small
compared to the two other forces.
At $t/\tau_{0}=500$,
before the tube reaches the uppermost CZ,
$z/H_{0}>-10$,
the horizontal flow is clearly driven
only by the gas pressure,
and, of course, the magnetic pressure gradient is zero.
Therefore,
we can conclude that
the HDF prior to the flux appearance
is caused by the pressure gradient.
This is consistent with
other numerical simulations
including thermal convection
\citep{che10}.
At $t/\tau_{0}=600$,
the shallow layer
is covered by the rising tube
and the gas pressure gradient reverses its sign.
Instead,
the magnetic pressure gradient
becomes dominant enough to drive the flow.

\begin{figure*}
  \sidecaption
  \includegraphics[width=12cm,clip]{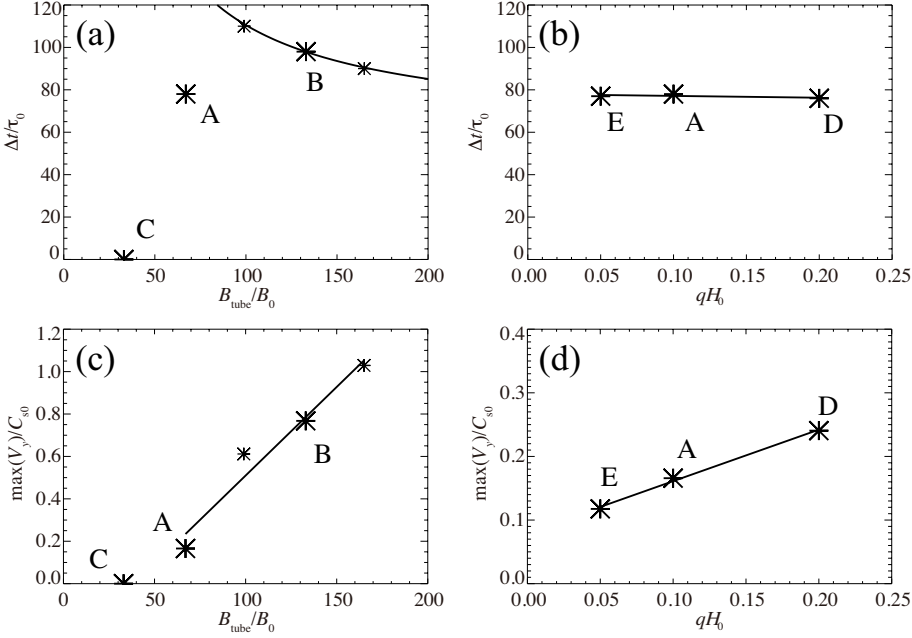}
  \caption{
    (a) Dependence of the HDF duration
    $\Delta t$
    on the field strength $B_{\rm tube}$,
    and (b) that on the twist $q$.
    (c) Dependence of 
    of the maximum HDF speed
    $\max{(V_{y})}$
    on the field strength $B_{\rm tube}$,
    and (d) that on the twist $q$.
    In Panel (a) and (c)
    we also plot other field strength cases,
    which are indicated by smaller asterisks.
    Solid lines are the fitted curves;
    (a) $\Delta t/\tau_{0}
    =5.08\times 10^{3} / (B_{\rm tube}/B_{0})
    +5.97\times 10$,
    (b) $\Delta t/\tau_{0}
    = -8.57 \times (qH_{0})
    + 7.80\times 10$,
    (c) $\max{(V_{y})}/C_{\rm s0}
    =8.35\times 10^{-3}
    \times (B_{\rm tube}/B_{0})
    -3.26\times 10^{-1}$,
    (d) $\max{(V_{y})}/C_{\rm s0}
    = 8.10\times 10^{-1}
    \times (qH_{0})
    +8.02 \times 10^{-2}$.
  }
  \label{fig:hdf}
\end{figure*}

\subsection{Dependence of the HDF
  \label{sec:hdf}}

In this subsection,
we show the dependence of the HDF
on the initial field strength
$B_{\rm tube}$
and on the twist $q$.
The investigated parameters are
the duration of the HDF
(from the HDF start to flux appearance),
$\Delta t/\tau_{0}$,
and the maximum HDF velocity,
$\max{(V_{y})}/C_{\rm s0}$,
during this time period.
Here we defined the start time of the HDF
as ``when the horizontal speed $V_{y}$
in the horizontal range
$-50\le y/H_{0}\le 50$,
averaged over
$6.5\le x/H_{0}\le 13.5$
and $-10\le z/H_{0}\le 0$,
exceeds $0.06C_{\rm s}(=0.5\ {\rm km\ s}^{-1})$''
and the flux appearance
as ``when the field strength $|B|$
in this range
exceeds $0.67B_{0}(=200\ {\rm G})$.''

Figs. \ref{fig:hdf}a and b
show the dependence
of the duration
on the field strength $B_{\rm tube}$
and the twist $q$.
Panel (a) is the comparison
among the different field strength cases.
A comparison of cases A and B,
the middle and stronger field tubes,
shows that
the time duration is longer
for the stronger field.
If other stronger cases are considered
(here we also plot two stronger tube cases
other than A, B, and C),
however,
it may be found that
the duration decreases
with field strength.
Thus,
we can divide these cases
into two groups:
stronger cases
that show a decreasing trend,
which is fitted
by a function of
$B_{\rm tube}^{-1}$,
and a middle case
that deviates
from the decreasing trend.
The weakest tube,
case C,
did not reach the surface.
That is why the duration is $0$
for case C.
In contrast,
in Panel (b),
the duration is almost constant
for the different twist cases.

Dependence of the maximum HDF speed,
$\max{(V_{\rm y})}$,
is shown in Figs. \ref{fig:hdf}c and d.
Panel (c) indicates
the positive linear correlation
with the field strength
$B_{\rm tube}$.
Again,
the speed of case B is
plotted as zero,
since it did not reach the surface.
Note that,
in Panel (a),
we found a gap
between the middle-field regime
and the stronger-field regime.
Thus
the linear fitting
in Panel (c)
might not reflect
the actual trend.
Nevertheless,
the maximum speed
basically increases
with field strength.
In Panel (d),
we can see that
the maximum HDF velocity
is clearly proportional
to the initial twist $q$.

\section{Analytic Explanation
  \label{sec:analytic}}

In this section,
the dependencies
of the rising speed
and of the HDF
on the physical parameters
obtained in Section \ref{sec:results}
are analytically explained.

\subsection{Rising Speed
  \label{sec:speed}}

In Section \ref{sec:general},
we found that
the rising speed
of the flux tube
is proportional
to the field strength $B_{\rm tube}$
and the dependence on the twist $q$
is significantly small.
The curvature is effective
for the flux tube with
the shortest wavelength $\lambda$.
Here we assume that
the rising speed in the CZ
is given as a terminal velocity
where the buoyancy of the tube
equals the aerodynamic drag
by the surrounding flow field
\citep{par75,mor96}
and the downward magnetic tension.
Buoyancy, dynamic drag,
and tension force
acting on a unit cross-sectional area
are written as
\begin{eqnarray}
  f_{\rm B}=\frac{B^{2}}{8\pi H_{\rm p}},
\end{eqnarray}
\begin{eqnarray}
  f_{\rm D} = C_{\rm D} \frac{\rho V_{z}^{2}}{\pi R_{\rm tube}},
\end{eqnarray}
and
\begin{eqnarray}
  f_{\rm T} = \frac{B^{2}}{4\pi R_{\rm c}},
\end{eqnarray}
respectively,
where $H_{\rm p} = H_{0}(T/T_{0})$ denotes
the local pressure scale height,
$V_{z}$ the tube's vertical speed,
$C_{\rm D}$ the drag coefficient
of order unity,
and $R_{\rm c}$ is
the radius of curvature.
The mechanical balance
$f_{\rm B}=f_{\rm D}+f_{\rm T}$
yields the terminal velocity
\begin{eqnarray}
  V_{\infty}^{2}
  =
  \frac{R_{\rm tube}B^{2}}{4C_{\rm D}\rho}
  \left(
    \frac{1}{2H_{\rm p}} - \frac{1}{R_{\rm c}}
  \right).
  \label{eq:terminal1}
\end{eqnarray}

\begin{figure}
  \centering
  \includegraphics[scale=1.,clip]{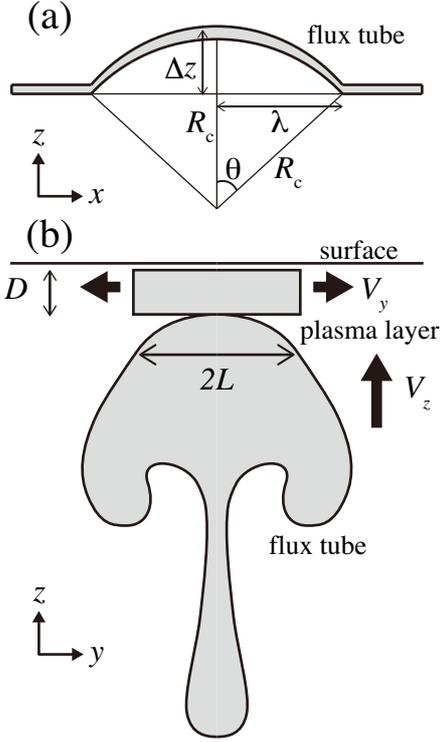}
  \caption{(a) Cross-section
    of the rising flux tube
    along the axis
    (in the $x-z$ plane).
    (b) Cross-section
    of the rising flux tube
    and the plasma layer
    ahead of the tube
    (in the $y-z$ plane).}
  \label{fig:model}
\end{figure}

First, let us discuss
the curvature effect.
In Equation (\ref{eq:terminal1}),
the tension force is negligible
for $R_{\rm c}\rightarrow \infty$,
while the tension becomes effective
when $R_{\rm c}\sim 2H_{\rm p}$.
The relationship between
the curvature radius $R_{\rm c}$
and the perturbation wavelength
$\lambda$ is illustrated
as Fig. \ref{fig:model}a.
Here, we write
the tube's height as
$\Delta z = z_{\rm top}(t)-z_{\rm tube}$.
From this figure,
we have
\begin{eqnarray}
  \left\{
    \begin{array}{l}
      R_{\rm c} - R_{\rm c}\cos{\theta} = \Delta z\\
      R_{\rm c}\sin{\theta} = \lambda
    \end{array}
    \right.,
\end{eqnarray}
which gives
\begin{eqnarray}
  \lambda = \sqrt{2R_{\rm c} \Delta z
    - (\Delta z)^{2}}.
\end{eqnarray}
Thus,
using the condition
$R_{\rm c}\sim 2H_{\rm p}$,
we obtain
the critical wavelength
for the tension to be effective:
\begin{eqnarray}
  \lambda_{\rm c}
  = \sqrt{4H_{\rm p} \Delta z
    - (\Delta z)^{2}}.
\end{eqnarray}
For instance,
when the tube is
halfway to the surface,
i.e., $z_{\rm top}/H_{0}=-50$
and thus $\Delta z/H_{0}=50$,
the local pressure scale height
at this depth
is $H_{\rm p}/H_{0}\sim 21$.
Therefore,
the critical wavelength
is evaluated to be
$\lambda_{\rm c}\sim 41.2H_{0}$,
and the flux tube with a wavelength
smaller than this value
will be resisted
by the tension force,
$\lambda\la \lambda_{\rm c}\sim 41.2H_{0}$. 
In Fig. \ref{fig:results}c,
we found that
only the tube with $\lambda=25H_{0}$
shows slower emergence
due to the effective curvature force,
which satisfies the condition
$\lambda\la \lambda_{\rm c}$.

Next, let us go on to
the dependencies
on the field strength and the twist,
by considering
$R_{\rm c}\rightarrow \infty$.
Now the equation of
the terminal velocity
(\ref{eq:terminal1})
reduces to
\begin{eqnarray}
  V_{\infty}
  &=& \sqrt{ \frac{R_{\rm tube}}{8C_{\rm D}H_{\rm p}\rho} }\,
  B_{\rm tube} \sqrt{1+q^{2}r^{2}}\,
  \exp{\left( -\frac{r^{2}}{R_{\rm tube}^{2}} \right)}
  \nonumber \\
  &\sim& \sqrt{ \frac{e^{-2}R_{\rm tube}}{8C_{\rm D}H_{\rm p}\rho} }\,
  B_{\rm tube} \sqrt{1+q^{2}R_{\rm tube}^{2}}.
  \label{eq:terminal2}
\end{eqnarray}
Here, in the first line
we use Equation (\ref{eq:tube})
and in the second line
we assume $r\sim R_{\rm tube}$.
From this equation,
we see that the rising velocity
is in simple proportion to
the initial field strength $B_{\rm tube}$
when $q$ is constant.
If we change $q$
with considering $R_{\rm tube}=5H_{0}$,
for $qH_{0}=[0.05, 0.1, 0.2]$,
the third term in the right-hand-side
of Equation (\ref{eq:terminal2}) gives
\begin{eqnarray}
  \sqrt{1+q^{2}R_{\rm tube}^{2}}=\left\{
    \begin{array}{l}
      1.03\\
      1.12\\
      1.4
    \end{array}
    \right..
\end{eqnarray}
That is,
the second term has
only a weak positive correlation
to the value of $q$.
Therefore,
the rising velocity
of the flux tube
is proportional to
the field strength,
while it is almost independent
on the initial twist.
The trend of the rising speed
found in Section \ref{sec:general}
is thus explained.

Note here that,
if we substitute $H_{\rm p}\sim 40H_{0}$,
$\rho\sim 275\rho_{0}$
(values at $z_{\rm tube}$),
$B_{\rm tube}=67B_{0}$,
$q=0.1/H_{0}$, and $R_{\rm tube}=5H_{0}$
(values for case A)
and assume $C_{\rm D}\sim 1$
in Equation (\ref{eq:terminal2}),
we obtain $V_{\infty}=0.21C_{\rm s0}$,
which is comparable to the simulation result
$\sim 0.17C_{\rm s0}$
(Fig. \ref{fig:results}).
This agreement indicates
that Equation (\ref{eq:terminal2})
is a rather reasonable estimation
of the tube's rising speed
\citep[see also][]{par75,mor96}.

\subsection{Dependence on the Twist}

We found
in Section \ref{sec:hdf}
that
when the twist $q$ is varied
while the field strength $B_{\rm tube}$
is kept constant,
the duration of the HDF
$\Delta t$
is almost constant
while the maximum horizontal speed
$\max{(V_{y})}$
is proportional to $q$.

This feature can be explained
by considering a simple model
illustrated as Fig. \ref{fig:model}b.
Here the flux tube with a head size of $2L$
is rising at $V_{z}$,
which pushes the plasma layer
with a thickness $D$.
The thickness $D$
is also described as
$D\sim |z_{\rm top}(t)|$,
where
$z_{\rm top}(t) = z_{\rm tube}
+ \int_{0}^{t} V_{z}(t') dt'$.
From the discussion in Section \ref{sec:speed},
$V_{z}$ and thus $D$
are independent of $q$,
which indicates that
the HDF duration $\Delta t\simeq D/V_{z}$
is also independent of $q$.

If we write
the outflow speed as $V_{y}$,
mass flux conservation
can be written as
\begin{eqnarray}
  V_{y}=\frac{V_{z}}{D}L.
  \label{eq:conservation}
\end{eqnarray}
Here, $V_{z}/D$ is
independent of $q$.
The head size of the tube $L$,
however,
depends on the twist $q$,
since the aerodynamic drag
peels away the tube's outer flux
and its amount depends on the twist.
The head size remains larger with $q$,
which results in the stronger HDF;
$V_{y}\propto L(q)$.
Thus the maximum speed,
$\max{(V_{y})}$,
will also depend on $q$.

It should be noted here
that $L$ and $q$
are not always linearly correlated.
According to \citet{mor96},
the boundary of the expanded tube
is well defined
by the equipartition surface,
where the kinetic energy density
equals the magnetic energy density
of the
azimuthal field:
\begin{eqnarray}
  \frac{1}{2}\rho V^{2}
  =\frac{B_{\phi}^{2}}{8\pi}.
  \label{eq:equi}
\end{eqnarray}
On the basis of an analogy
from Equation (\ref{eq:tube}),
the profile of the expanded tube
at the equipartition surface,
where the radial distance is $r_{1}$
(the subscript ``1'' indicates
the expanded tube),
is assumed to be written as
\begin{eqnarray}
  \left\{
    \begin{array}{lll}
      B_{x}(r_{1}) &=&
      B_{\rm tube 1}
      \exp{\left( -\displaystyle\frac{r_{1}^{2}}{R_{\rm tube 1}^{2}} \right)}\\
      B_{\phi}(r_{1}) &=& qr_{1}B_{x}(r_{1})
    \end{array}
    \right.,
\end{eqnarray}
where $B_{\rm tube 1}$ and $R_{\rm tube 1}$
are the axial field and the typical radius,
respectively.
Then,
Equation (\ref{eq:equi}) reduces to
\begin{eqnarray}
  \frac{r_{1}}{R_{\rm tube1 }}
    \exp{\left( -\frac{r_{1}^{2}}{R_{\rm tube 1}^{2}} \right)}
  = \left( \frac{4\pi\rho V^{2}}{B_{\rm tube1}^{2}} \right)^{1/2}
  \frac{1}{qR_{\rm tube1}}.
  \label{eq:expand}
\end{eqnarray}
Here we assume that
$V$ is more or less approximate to $V_{z}$
and thus $V$ does not depend on $q$.
Other values of $B_{\rm tube 1}$,
$R_{\rm tube 1}$, and $\rho$
are also assumed to be constant
for different $q$.
Here we introduce
the notation
$\mu \equiv r_{1}/R_{\rm tube 1}$.
Then Equation (\ref{eq:expand})
reduces to
\begin{eqnarray}
  \mu \exp{(- \mu^{2})} = 1 / \hat{q},
\end{eqnarray}
or,
\begin{eqnarray}
  \mu^{2} - \ln{\mu} = \ln{ \hat{q} },
\end{eqnarray}
where
$\hat{q}\equiv qR_{\rm tube1} [ B_{\rm tube1}^{2}/(4\pi\rho V^{2})]^{1/2}$.
For a larger radial distance,
$r_{1} \gg R_{\rm tube 1}$,
i.e., $\mu \gg 1$,
\begin{eqnarray}
  \mu^{2} \sim \ln{ \hat{q} }.
\end{eqnarray}
Then we obtain
\begin{eqnarray}
  r_{1} \sim R_{\rm tube1}
  \left[
    \ln{(qR_{\rm tube1})}
    +\frac{1}{2}
    \ln{ \left( \frac{B_{\rm tube1}^{2}}{4\pi\rho V^{2}} \right)}
  \right]^{1/2}.
\end{eqnarray}
Therefore,
the effective size of the expanded tube $L\sim r_{1}$
is at least positively correlated to $q$,
but not in a linear manner.

\subsection{Dependence on the Field Strength}

When the field strength
at the tube's axis $B_{\rm tube}$
is varied
while the twist $q$
is fixed,
the maximum HDF speed,
$\max{(V_{y})}$,
is found to be roughly proportional
to the field strength.
From Equation (\ref{eq:conservation}),
if we assume $L/D$ is constant,
the horizontal speed $V_{y}$
and thus the maximum speed $\max{(V_{y})}$
are proportional
to the field strength $B_{\rm tube}$.

As for the HDF duration $\Delta t$,
however,
Fig. \ref{fig:hdf}a
clearly shows
two regimes:
stronger field cases
that show a decreasing trend,
and a middle case
that deviates from this trend.
Thus we should take into account
the difference between these regimes.

First,
let us focus on
the stronger field regime.
Since the rising speed
is proportional to the field strength,
stronger tubes emerge faster.
In this case,
the accumulated plasma
ahead of the tube
does not drain down so much
because of the short emergence period,
and thus the thickness of the plasma layer
becomes almost the same
for these cases.
That is,
the thickness of the layer $D$
is constant and is independent
of the rising speed $V_{z}$.
Since the rising speed
is proportional to the field strength
$B_{\rm tube}$,
we have
\begin{eqnarray}
  \Delta t &\simeq& D / V_{z}
   \propto 1 / B_{\rm tube}.
\end{eqnarray}
Hence,
the HDF duration
is inversely proportional
to the field strength,
which explains the trend
in the stronger field regime of
the fitted inverse function
in Fig. \ref{fig:hdf}a.

As for the middle strength case,
the emergence takes longer
and thus the drainage
of the accumulated plasma
becomes more effective,
resulting in the much thinner layer $D$
compared to the rising speed $V_{z}$.
Therefore,
the HDF duration
$\Delta t\simeq D/V_{z}$
becomes shorter
and thus deviates
from the inverse trend
in the stronger field cases.

\section{Summary
  \label{sec:summary}}

In this parametric survey,
we vary the axial field strength,
twist, and perturbation wavelength
of the initial flux tube.
As a result,
we found the following features.

The rising speed in the CZ
strongly depends on the initial field strength
but its
correlation with the twist is weak.
The emergence is resisted
by the curvature force
only in the case when
the perturbation wavelength
is shortest.
According to the analytic study,
the rising rate (terminal velocity)
is written as
$V_{\infty}\propto B_{\rm tube}\sqrt{1+q^{2}R_{\rm tube}^{2}}$,
which indicates a strong dependence
on the field strength
and a weak correlation with the twist.

As the flux tube approaches the surface,
the accumulated plasma ahead of the tube
escapes horizontally around the surface layer,
which is called the HDF.
The driver of the HDF is
found to be the pressure gradient.

When the field strength increases,
the maximum HDF speed becomes higher,
because the rising speed
mainly depends on the field strength.
The HDF duration is divided into two groups.
For the stronger tube regime
($B_{\rm tube}\ga 100B_{0}$),
the duration is
in simple inverse proportion
to the field strength,
while the weaker field regime
($B_{\rm tube}\la 100B_{0}$)
deviates from the trend
in the stronger tube regime
because the fluid draining
is more effective.

The duration of the HDF
is found to have
no relation with the tube's twist,
since the rising speed
is independent of the twist.
However, the maximum HDF speed
shows a positive correlation with the twist.
This feature is explained
by considering the head size of the main tube
that remains after the aerodynamic drag
peels away the tube's outer field.
The head size remains larger
with the twist,
which results in the stronger HDF.

If we apply the above dependencies
of the HDF
to the actual observations,
we may be able to
obtain information
on the magnetic field
in the subsurface layer,
which we cannot observe optically.

\begin{acknowledgements}
  The authors would like to thank
  the anonymous referee
  for improving the paper.
  S.T. is supported by
  Grant-in-Aid for JSPS Fellows.
  Numerical computations were
  carried out on NEC SX-9 and Cray XT4
  at the Center for Computational Astrophysics, CfCA,
  of the National Astronomical Observatory of Japan.
  We are grateful to the GCOE program instructors
  of the University of Tokyo
  for proofreading/editing assistance.
\end{acknowledgements}

\bibliographystyle{aa}
\bibliography{reference}

\begin{thebibliography}{15}
\expandafter\ifx\csname natexlab\endcsname\relax\def\natexlab#1{#1}\fi

\bibitem[{{Archontis} {et~al.}(2004){Archontis}, {Moreno-Insertis},
  {Galsgaard}, {Hood}, \& {O'Shea}}]{arc04}
{Archontis}, V., {Moreno-Insertis}, F., {Galsgaard}, K., {Hood}, A., \&
  {O'Shea}, E. 2004, \aap, 426, 1047

\bibitem[{{Cheung} {et~al.}(2010){Cheung}, {Rempel}, {Title}, \&
  {Sch{\"u}ssler}}]{che10}
{Cheung}, M.~C.~M., {Rempel}, M., {Title}, A.~M., \& {Sch{\"u}ssler}, M. 2010,
  \apj, 720, 233

\bibitem[{{Emonet} \& {Moreno-Insertis}(1998)}]{emo98}
{Emonet}, T. \& {Moreno-Insertis}, F. 1998, \apj, 492, 804

\bibitem[{{Fan}(2001)}]{fan01}
{Fan}, Y. 2001, \apjl, 554, L111

\bibitem[{{Linton} {et~al.}(1996){Linton}, {Longcope}, \& {Fisher}}]{lin96}
{Linton}, M.~G., {Longcope}, D.~W., \& {Fisher}, G.~H. 1996, \apj, 469, 954

\bibitem[{{Magara}(2001)}]{mag01}
{Magara}, T. 2001, \apj, 549, 608

\bibitem[{{Moreno-Insertis} \& {Emonet}(1996)}]{mor96}
{Moreno-Insertis}, F. \& {Emonet}, T. 1996, \apjl, 472, L53

\bibitem[{{Murray} {et~al.}(2006){Murray}, {Hood}, {Moreno-Insertis},
  {Galsgaard}, \& {Archontis}}]{mur06}
{Murray}, M.~J., {Hood}, A.~W., {Moreno-Insertis}, F., {Galsgaard}, K., \&
  {Archontis}, V. 2006, \aap, 460, 909

\bibitem[{{Parker}(1975)}]{par75}
{Parker}, E.~N. 1975, \apj, 198, 205

\bibitem[{{Sch\"{u}ssler}(1979)}]{sch79}
{Sch\"{u}ssler}, M. 1979, \aap, 71, 79

\bibitem[{{Shibata} {et~al.}(1989){Shibata}, {Tajima}, {Steinolfson}, \&
  {Matsumoto}}]{shi89}
{Shibata}, K., {Tajima}, T., {Steinolfson}, R.~S., \& {Matsumoto}, R. 1989,
  \apj, 345, 584

\bibitem[{{Toriumi} {et~al.}(2012){Toriumi}, {Hayashi}, \& {Yokoyama}}]{tor12b}
{Toriumi}, S., {Hayashi}, K., \& {Yokoyama}, T. 2012, \apj, 751, 154

\bibitem[{{Toriumi} \& {Yokoyama}(2010)}]{tor10}
{Toriumi}, S. \& {Yokoyama}, T. 2010, \apj, 714, 505

\bibitem[{{Toriumi} \& {Yokoyama}(2011)}]{tor11}
{Toriumi}, S. \& {Yokoyama}, T. 2011, \apj, 735, 126

\bibitem[{{Toriumi} \& {Yokoyama}(2012)}]{tor12a}
{Toriumi}, S. \& {Yokoyama}, T. 2012, \aap, 539, A22

\end{thebibliography}

\end{document}